# Generation Efficiency of the Second Harmonic Inhomogeneous Component

V. Roppo[1,2], F. Raineri[3,4], C. Cojocaru[1], R. Raj[3], J. Trull[1], I. Sagnes[3], R. Vilaseca[1], M. Scalora[3]

[1]*Universitat Politècnica de Catalunya, Departament de Física i Enginyeria Nuclear, Colom 11, 08222 Terrassa, Spain*

[2]*Charles M. Bowden Research Facility, US Army RDECOM, Redstone Arsenal AL 35803, USA*

[3]*Laboratoire de Photonique et de Nanostructures, Route de Nozay 91460 Marcoussis, France*

[4]*Université Paris Diderot, Paris, France*

**Abstract**

In this letter we experimentally demonstrate second harmonic conversion in the opaque region of a GaAs cavity with efficiencies of the order of 0.1% at 612nm, using 3ps pump pulses having peak intensities of order of 10MW/cm$^2$. We show that the conversion efficiency of the inhomogeneous, phase-locked second harmonic component is a quadratic function of the cavity factor Q.

**Introduction**

Second harmonic generation (SHG) [1, 2, 3] is among the most used and studied nonlinear phenomena. When a fundamental (FF) electromagnetic field impinges on a quadratic nonlinear material the resulting polarization is a quadratic function of the field. The second harmonic (SH) signal has a carrier frequency that doubles the FF one. Even though the general solution of a problem that includes dispersion and absorption is rather complicated, it is still possible to derive approximate solutions that give a fairly good idea about the shape and behavior of the generated SH [1]. The general solution comes from an inhomogeneous set of differential equations, and consists of two parts: the first portion is the solution of the homogeneous (HOM) set of equations, while the second part is a particular solution of the inhomogeneous (INH) set of equations. The former (HOM-SH) has a well known expression and its behavior has been intensely investigated, especially under phase matching (PM) conditions [4-6]. The latter (INH-SH) does not have an exact or unique form. Since the inhomogeneous solution manifests itself mostly under phase mismatched conditions, its conversion efficiencies are generally very low compared to phase matched cases. As a result, in the past the study of this SH component has generated very little interest.

During the last few decades only a small number of studies have reported on this twofold nature of the generated SH signal, an aspect that gives rise to clearly identifiable Maker fringes in the steady state[7, 8], and to a double-peaked SH signal shape [9-11]. Even if these studies were carried out in a strong phase mismatched regime, in general, most efforts focused mostly towards the study of the phase matched SHG to ensure maximum conversion efficiency. As became clear later [12], in uniform media only the HOM-SH component can be enhanced. This situation makes the INH-SH component even more difficult to observe.

This relative lack of knowledge regarding the dynamics of pulses generated in the mismatched regime and the availability of laser sources with ever decreasing pulse durations called for a systematic investigation of the INH-SH component [12, 13]. Since it is not an easy task to ascertain its properties from the implicit form of the analytical solution, one has to rely on numerical studies. In refs.[12] and [13] the dynamics of the INH-SH component is discussed in some details. This component of the SH signal is *always* present during propagation and *always* travels bounded to the FF with the same energy velocity. Once generated within the first few coherence lengths as the pump pulse traverse the entry interface, the energy of the INH-SH signal clamps with no further energy exchange with the pump pulse. Even in a general phase mismatched working condition, the carrier wave vector of the trapped INH-SH pulse is always twice that of the FF, leading to a phase locking phenomenon. A simple way to sum up all these characteristics is to state that the INH-SH component reads the same effective complex index of refraction as the FF field. Thus, ultrashort pulse propagation in long crystals leads to the clear identification of the INH-SH preceded by a rapid walk-off of the HOM component, which propagates in the medium decoupled from the pump field, and experiences the nominal material dispersion expected at the SH frequency.

Experimental verification of these phenomena has been reported [13, 14, 15]. In particular, in references [13] and [14] the delay that appears due to the different effective indices of refraction experienced by the two beams in collinear propagation and the different refraction angles was measured for both INH-SH and HOM-SH components. In ref. [15] it was shown that the effective imaginary part of the index of refraction of the locked INH-SH pulse was identical to that of the pump. A FF beam tuned in a transparency region (1300nm) of a GaAs slab 500μm thick generated second (650nm) and third (433nm) harmonic signals deep in the absorption region. The absorption band edge for this material is located at ~900nm. As expected, the HOM-SH is quickly absorbed while the INH-SH component traverses the entire sample without losses. This effect can be understood as *inhibition of absorption* at the harmonic wavelengths. This result presents us with many new opportunities for at least two reasons. On one hand these findings extend the useful spectral range for harmonic generation, pushing it well into the UV and X-UV frequencies. On the other hand, absorption removes the homogeneous component, allowing us to focus on the study of the INH-SH component only.

We pose the following question: is it possible to enhance SH in the phase mismatched regime using ordinary laser sources and materials, up to a level that could rival harmonic generation from phase-matched or quasi phase matched materials? To take advantage of the inhibition of absorption of the INH-SH component and find an affirmative answer to this question it is necessary to improve the nonlinear conversion efficiencies.. As reported in references [7-13, 15], using pump intensities of order 1GW/cm$^2$, conversion efficiencies from a single interface of GaAs or similar material is of order 10$^{-10}$ Presently we are not aware of analytical tools that might be used to infer the explicit expression of conversion efficiency for the phase locked component.

In reference [16] the behavior of the INH-SH in a cavity environment was theoretically predicted. The GaAs cavity was designed to be resonant at the FF frequency (1300nm), with consequent obvious detuning for all harmonics due to material dispersion, absorption notwithstanding. The results demonstrated that the HOM-SH is mostly absorbed, while the INH-SH becomes localized inside the cavity with an effective index that matched the FF index of refraction. This localization, together with the localization of the FF field and the resulting good overlap between the fields, leads to improvements of the SH conversion efficiency. The effect was later demonstrated experimentally in reference [17], where a simple cavity with a quality factor of order 10 sufficed to improve the efficiency by at least two orders of magnitude, up to ~10$^{-7}$. A general estimation, $SH_{eff} \propto Q^2$, was proposed based on Fermi's golden rule. In what follows we present results that confirm that this simple relation is a useful and practical tool to predict conversion efficiencies of the INH-SH component.

In our experiment we used four different samples made of a 350nm thick GaAs layers sandwiched between different sets of distributed Brag mirrors (inset Fig.1). The samples were designed with 4, 6, 8 and 10 pairs of SiO$_2$(120nm)/Si$_3$N$_4$(210nm) layers on each side, respectively. Each of these samples has a resonance centered at approximately the same wavelength, but having different width due to their different quality factors. Experimentally, the resonance shape and width may be ascertained by illuminating the samples with an ultrashort femtosecond pulse. The measured resonances are shown in Fig.1. The resonances are all centered near λ=1225nm and have widths (FHMW) of 9.3nm, 2.6nm, 1.2nm, and 0.32nm, respectively. The corresponding cavity Q factors, calculated as Q=λ/Δλ, are 125, 350, 1200 and 3830.

The difference between this kind of dielectric sandwich and the free standing GaAs etalon discussed in reference [16] is that in this case the internal field is amplified by hundreds of times relative to the incident field. The linear properties of the stack are such that it is transparent at the FF wavelength and allows it to become localized, but it remains opaque for the SH (the transmission at 612nm is less than 1%). This guaranties that the observed SH signal consists of INH-SH component only.

The FF pulse centered at λ=1225 nm was provided by a tunable OPA and was slightly focused down to a spot of about 500μm onto the sample. Its duration was fixed at 3ps and the intensity incident onto the sample was of the order of 10MW/cm$^2$. The transmitted signals at the FF wavelength were collected with a cooled AlGaAs camera, and the ones at the SH frequency were recorded with a Si detector. The sample is rotated 10º degree to take advantage of the non vanishing $\chi^{(2)}$ in that direction. The experimental set up is sketched in the inset of Fig.3.

The SH efficiency was normalized to the effective FF energy that finds its way inside the cavity. To account for the spectral width of the pulse, for each sample we calculated the overlap between the FF and the cavity spectra. In short, after we put the sample in place in the experimental set-up the laser source was shifted to shorter duration (around 80fs). This broad spectrum pulse is able to scan the effective transmission characteristics of the sample. Multiplying this spectrum for the spectrum of the 3ps pulse used during the experiment we can estimate how much FF energy has actually penetrated inside the cavity.

For each sample we first measured the behavior of the SH output power versus the input FF power (Fig.2). As shown in the figure, the SH follows a regular quadratic pattern. This is an indication that two-photon absorption and other non-linear effects are not affecting the generation process in an appreciable measure. Notice how the FF power range becomes smaller for higher Q cavities due to the smaller bandwidth of the resonance. To operate in a safe region for all sample, we chose to extrapolate the SH efficiencies corresponding to 0.15MW/cm$^2$ of FF intensity. This corresponds more or less to the last experimental point for the 10-mirrors pairs sample in Fig.2.

Finally, in Fig.3 we report for each sample the SH conversion efficiency vs. cavity factor Q. The circles show the experimental points and the solid line is a quadratic fit of the data. The first circle marks the bulk emission. The results are thus in good agreement with the general estimate made in reference [17]. The figure shows that it is possible to achieve efficiency of the order of 0.1% with external pump peak intensities of order 10MW/cm$^2$, which for 3ps incident pulses leads to 0.15MW/cm$^2$ inside the cavity. This means that there is real potential for pump depletion with relatively low intensities (tens of MW/cm$^2$), provided the incident pulse resolves the resonance. In this particular situation the FF peak power was kept conservatively low to avoid two-photon absorption at the pump frequency, which can be considerable in GaAs at these wavelengths. Beyond this, the INH-SH behavior is completely general and possible devices could be designed by choosing more appropriate materials, such as polymers, to outperform GaAs given the potential for much higher nonlinear coefficients [18]. Finally, these results are easily extended to third harmonic generation, as shown in references [15-17]. In this particular case the cavity was not optimized for third harmonic generation, which happened to be tuned inside the band gap of the stack.

In conclusion, we experimentally demonstrated SH conversion efficiencies that approach 0.1% in an opaque GaAs cavity, using incident pump intensities of order 10MW/cm$^2$. We have also shown that the conversion efficiency of the inhomogeneous, phase locked SH component is a quadratic function of cavity-Q, namely the enhancement factor of the FF field inside the cavity. These results show that high conversion efficiencies can be obtained in phase mismatched materials in ways that may also be practical. In addition, with the right materials it becomes possible to exploit new wavelength ranges, e.g. UV-XUV, using GaP or GaN, for example.

**Figures and figures captions**

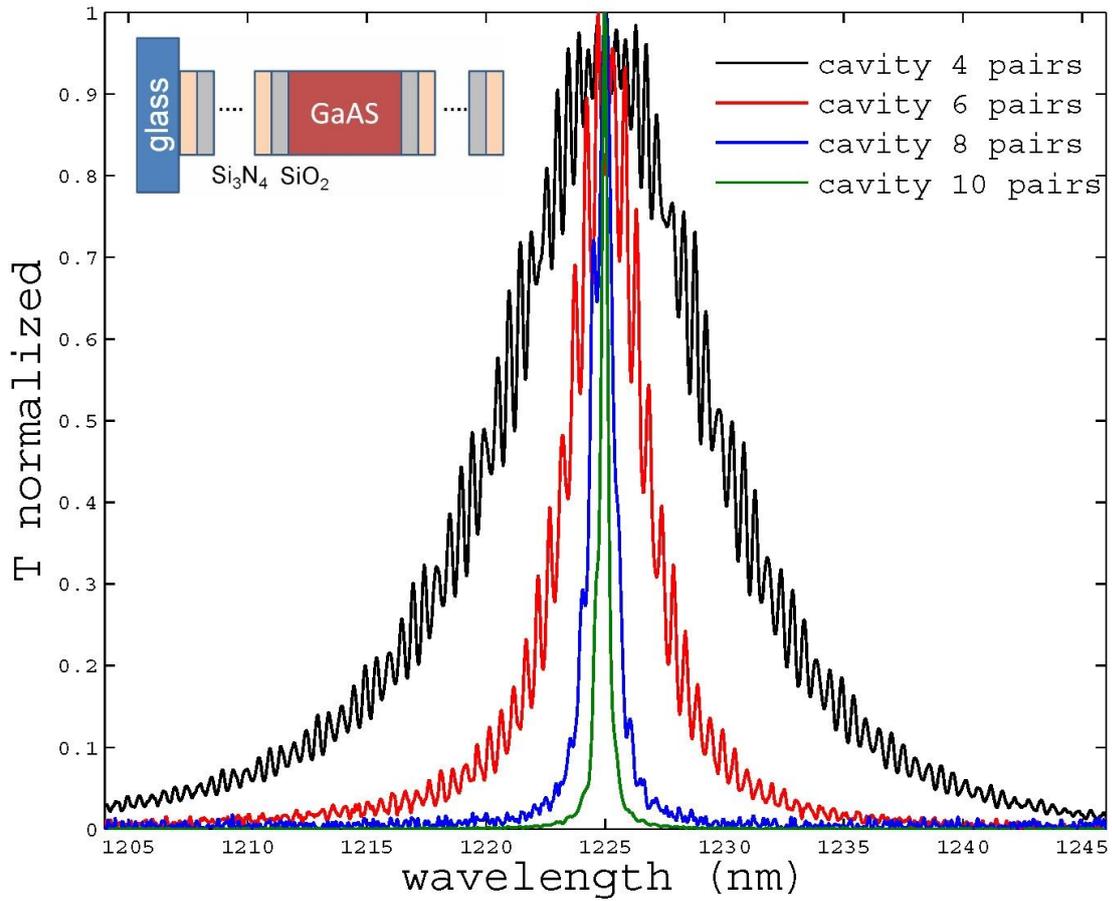

**Figure 1**: Samples transmission spectra. The transmission curves are obtained from the transmission spectra of an ultrashort femtosecond laser. The fast oscillations are due to the multiple reflections inside the 1mm glass substrate. The transmittance scale is normalized for graphical reasons. The peaks reach 0.74, 0.34, 0.26 and 0.06 for the 10, 8, 6, and 4 period mirror samples, respectively. As a reference, the FF field spectrum is around 2nm FWHM. Inset: sample scheme.

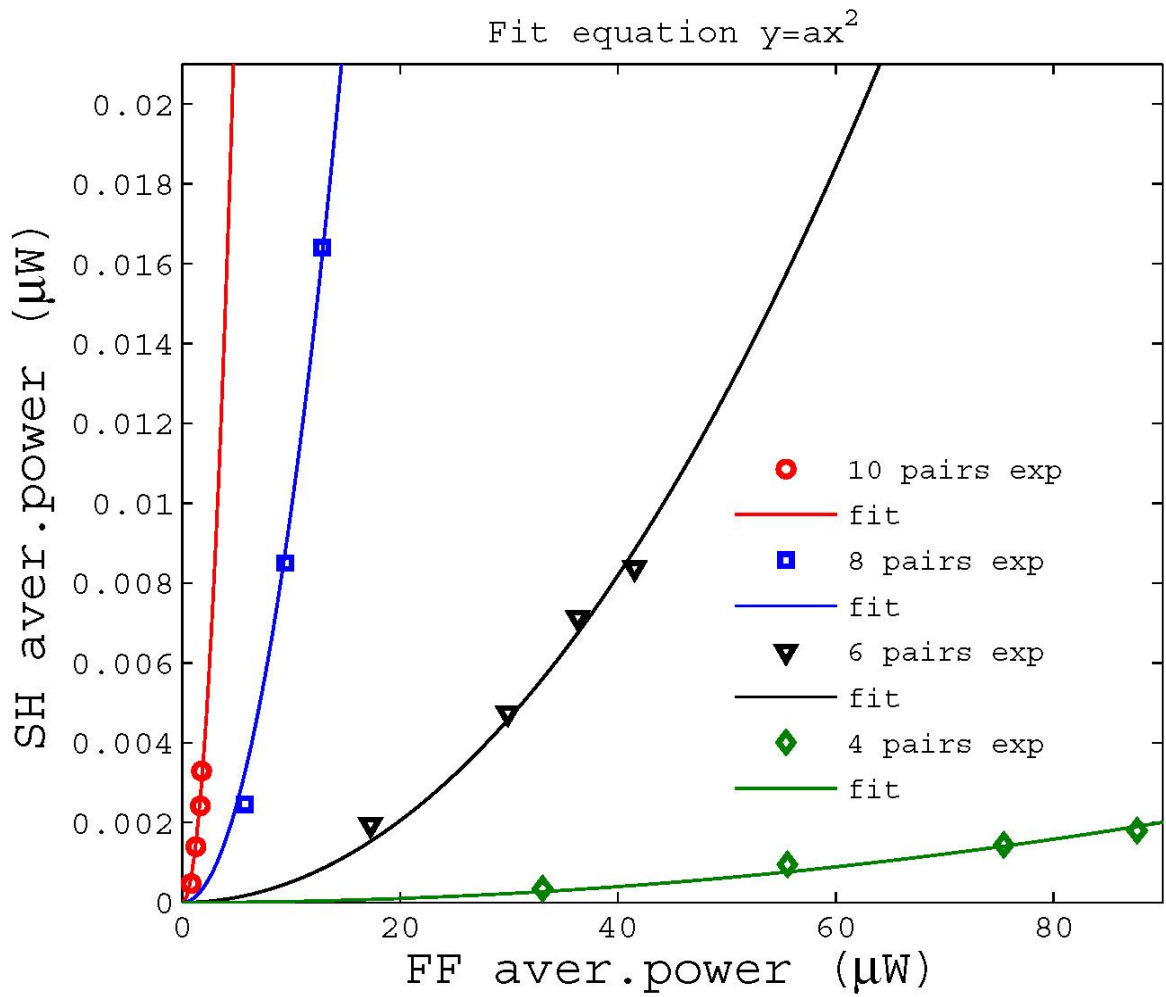

**Figure 2**: SH average power versus FF average power for all samples. The continuous lines are the quadratic fits of the experimental data (circles). The laser output power is always in the same range but the effective powers inside the cavities depend on resonance shape.

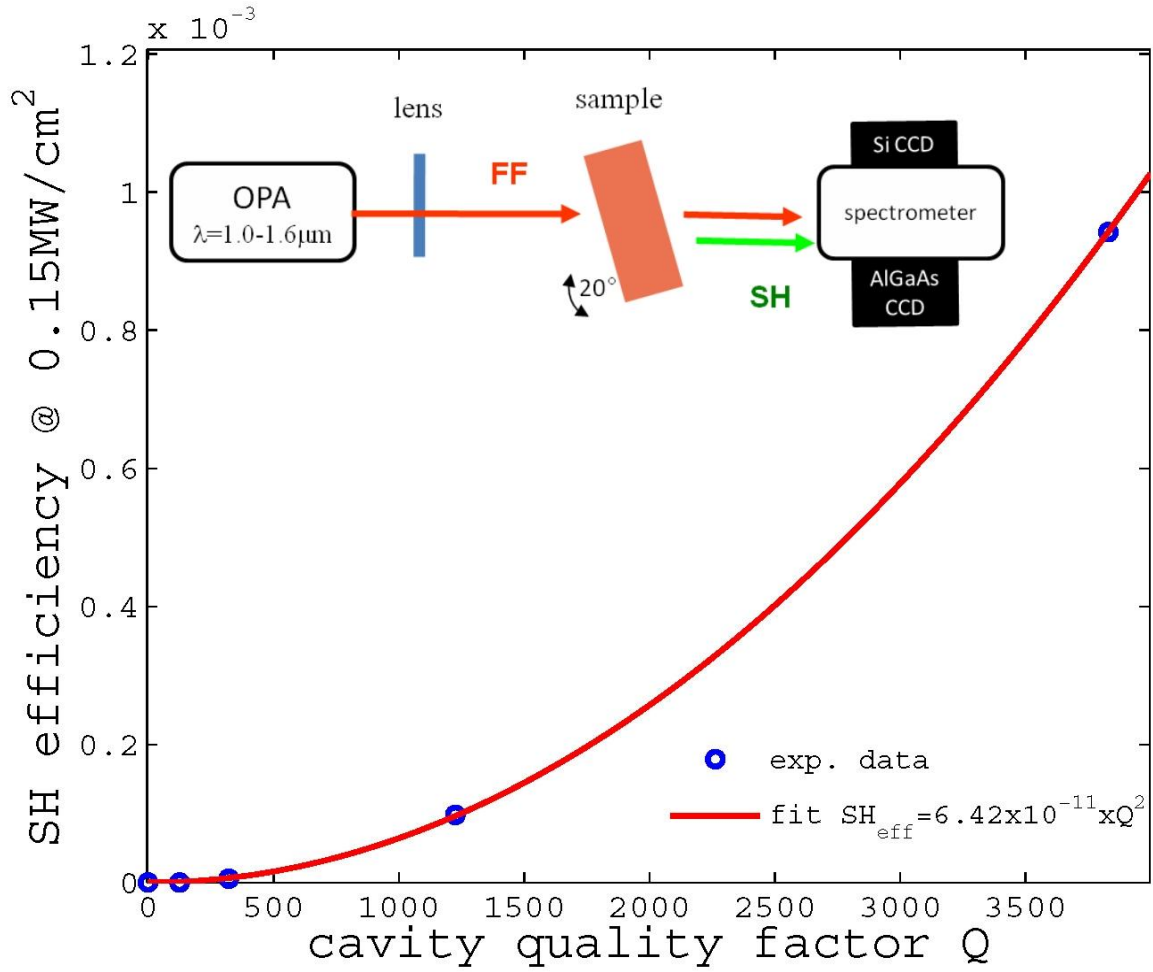

**Figure 3**: Second harmonic efficiency versus cavity Q factor, experimental data and quadratic fit. Due to the absorption regime at the harmonic wavelength, the SH is composed only by the inhomogeneous component. Inset: experimental set up.